\documentclass[osajnl,preprint,showpacs]{revtex4}
\DeclareRobustCommand{\baselinestretch{2}} 

\begin{document}

\title{All-optical switching and amplification of discrete vector solitons
in nonlinear cubic birefringent waveguide arrays}

\author{Rodrigo A. Vicencio}
\author{Mario I. Molina}
\address{Departamento de F\'{\i}sica, Facultad de Ciencias,
Universidad de Chile, Casilla 653, Santiago, Chile}
\author{Yuri S. Kivshar}
\address{Nonlinear Physics Centre, Research School of Physical
Sciences and Engineering, Australian National University, Canberra
ACT 0200, Australia}

\begin{abstract}We study all-optical switching based on the dynamical properties
of discrete vector solitons in waveguide arrays. We employ the
concept of the polarization mode instability and demonstrate
simultaneous switching and amplification of a weak signal by a
strong pump of the other polarization.
\end{abstract}

\ocis{190.0190, 190.4370, 190.5530.}

\maketitle

\section{Introduction}

Discrete spatial solitons are usually introduced as spatially
localized nonlinear modes of weakly coupled optical
waveguides~\cite{yuriagra}, and they have recently been observed
experimentally in different nonlinear systems~\cite{review}. A
standard theoretical approach in the study of discrete spatial
optical solitons is based on the derivation  of an effective
discrete nonlinear model and the analysis of its stationary
localized solutions---discrete localized modes~\cite{physto}.

Many studies analyzed both propagation and steering of {\em
scalar} discrete solitons in nonlinear waveguide arrays, including
trapping, reflection, and refraction of an incoming discrete
soliton in an array with defects~\cite{kroli}. For waveguide
arrays with quadratic nonlinear response, Pertsch {\em et
al.}~\cite{X2} suggested an optical switching scheme where a
low-power diffractionless beam is coupled parametrically to a
pump, resulting in generation of a strong idler beam. Recently, we
suggested~\cite{olours} to control multi-port switching of
discrete solitons in waveguide arrays by engineering the coupling
between the neighboring waveguides: this induces a change of the
dynamic properties of the array through the modification of the
effective Peierls-Nabarro (PN) potential , a nonlinear
discreteness-induced potential that is responsible for the
transverse dynamics of discrete solitons in a waveguide array.

In  materials with a nonlinear cubic response, we expect a very
rich nonlinear dynamics and coupling between different modes. In
particular, coupling between the waves of two orthogonal
polarizations can result in the formation of {\em a vector
soliton}. Vector solitons can be employed for different schemes of
all-optical switching schemes, e.g. by employing collisions
between orthogonally polarized solitons without~\cite{coll1} or
with~\cite{coll2} four-wave mixing (FWM) effects. The FMW effect
is responsible for the energy exchange between the two (TE and TM)
polarizations. In planar waveguides, the TE mode is known as
`slow', while the TM mode-- as `fast', because the TE mode has
larger propagation constant. Interaction between the TE and TM
modes is known to lead to polarization mode
instability~\cite{polins}, so that the TM mode is always unstable
with respect to the TE mode. This implies that the fast mode
transfers the energy to the slow mode.

The theory of discrete vector solitons developed so far does not
include the analysis of the FWM effects (as an example, see Ref.
\onlinecite{disvec}). However, the first experimental studies of
the vectorial interactions and discrete vector solitons in
waveguide arrays~\cite{vecexp} due to coupling between two
orthogonally polarized modes, suggest the importance of the FWM
effects demonstrating that the initial phase between the TE and TM
modes defines the energy exchange between the modes.

In this Letter, we study numerically the dynamics and switching
properties of discrete vector solitons employing the concept of
polarization mode instability. We demonstrate an effective
switching, control, and amplification of a weak signal by a strong
pump of the orthogonal polarization that can be useful for
multi-port all-optical switching.

In order to analyze the polarization effects for discrete
solitons, we consider a model of  birefringent cubic nonlinear
waveguide arrays recently fabricated experimentally~\cite{vecexp},
described by the coupled-mode theory within the slowly varying
envelope approximation:
\begin{eqnarray}
-i \frac{d a_{n}}{d z}=V_n^{+}\{a_n\} +|a_{n}|^{2}a_{n}+
A |b_{n}|^{2}a_{n}+B b_{n}^{2}a_{n}^{*} \nonumber\\
-i \frac{d b_{n}}{d z}= V_n^{-}\{b_n\} + |b_{n}|^{2}b_{n}+ A
|a_{n}|^{2}b_{n}+B a_{n}^{2}b_{n}^{*}, \label{01}
\end{eqnarray}
where $V_n^{\pm} \{a_n\} = \pm a_{n}+V(a_{n+1}+a_{n-1})$, $a_{n}$
and $b_{n}$ are the normalized envelopes of the TE- and
TM-polarized components of the electric field, respectively, $z$
stands for the propagation distance, $V$ is the coupling parameter
assumed to be the same for both polarizations (both modes have the
similar transverse extensions). Nonlinear coefficients $A$ and $B$
characterize the cross-phase-modulation and the FWM effects
(weighted with the self-focusing term), respectively. In our
study, we use the experimental parameters of this model as defined
in Ref.~\onlinecite{vecexp}, i.e. $A=1$, $B=0.5$, and $V=0.92$.
The normalized power,
\begin{equation}
P = P_a + P_b = \sum_{n} (|a_{n}|^{2}+|b_{n}|^{2}),
\end{equation}
and Hamiltonian, $H = -\sum_{n}(H^a_n + H_n^b + H_n^{ab})$, where
\begin{eqnarray}
H_n^{a,b} &=& \pm |a_{n}|^{2}+V(a_{n}^{*}a_{n+1}+
a_{n}a_{n+1}^{*})
+\frac{1}{2}|a_{n}|^{4}, \nonumber \\
H_n^{ab} &=&
|a_{n}|^{2}|b_{n}|^{2}+\frac{B}{2}(b_{n}^{2}a_{n}^{*2}+
a_{n}^{2}b_{n}^{*2}) \label{}
\end{eqnarray}
are conserved in the dynamics, and they both play an important
role for checking numerical accuracy and for estimation of the
realistic power of the switching-amplification process (e.g.,
$P_{real}\approx 56\times P \ [W]$ in the AlGaAs waveguide array
used in Ref.~\onlinecite{vecexp}).

An important step of our analysis is a choice of the appropriate
initial profiles for both modes. Since the exact solution to Eq.
(\ref{01}) is not known, we select a truncated sech-like
profile~\cite{olours}:
\begin{eqnarray}
a_n(0) = a_0\ {\rm sech} [a_0(n-n_{ca})/\sqrt{2}] \
e^{-ik_{a}(n-n_{ca})}\ e^{i \phi_a}\nonumber\\
b_n(0) = b_0\ {\rm sech} [b_0(n-n_{cb})/\sqrt{2}] \
e^{-ik_{b}(n-n_{cb})}\ e^{i \phi_b},\label{04}
\end{eqnarray}
for $n-n_{ca} = n-n_{cb}=0, \pm1$, and $a_n(0)=b_n(0)=0$,
otherwise. Parameters $a_0$ and $b_0$ are the initial amplitudes,
$k_{a}$ and $k_{b}$ are the initial kicks (transverse angles),
$n_{ca}$ and $n_{cb}$ are the initial center positions, and
$\phi_a$ and $\phi_b$ are the input phases, for both TE and TM
modes,  respectively. As has been verified earlier~\cite{olours},
this ansatz provides a very good agreement in the scalar case,
when just one mode propagates in the array. A specific choice of
the initial input is not a restriction to our analysis, because a
discrete soliton is a self-adjusting mode, and an energy excess is
emitted in the form of radiation modes~\cite{polins}. We have
carried out numerical simulations with other profiles, and the
dynamic behavior was found to be similar.

In the polarization-induced dynamics, a fast mode is unstable with
respect to its transformation into a slow
mode~\cite{polins,vecexp} Numerically, we observe that we can
invert this polarization instability with an appropriate use of
the initial phase shift. When the initial phase shift is $0$ or
$\pi$ (linearly polarized), the TM mode always transfers a part of
its energy to the TE mode.  On the other hand, if the initial
phase shift is $\pi /2$ (elliptically polarized), we observe a
reversed effect. Thus, we can control the energy transfer by
choosing the initial phase shift between both modes at the input.
In some cases, e.g. when we consider large propagation distances,
the energy exchange can be complete. In this paper, we study the
linearly polarized case only, because the maximum power gain for
the TE mode, $(P_{a}(z_{\rm max}) - P_{a}(0))/P_{a}(0)$, observed
in this situation ($\sim 6000\%$) is higher than for the
elliptically polarized case ($\sim 100\%$). This asymmetry in the
polarization instability is a very interesting feature, but here
we are interested only in an efficient all-optical switching
scheme based on this effect.

In our numerical simulations, we take $z_{\rm max} = 50$ and an
array of $110$ waveguides. We made sweeps of the system parameters
and observe the TE gain. On a phase diagram in Fig.~\ref{fig1}, we
show the TE (signal) power gain vs. the TM (pump) mode input angle
$k_{b}$ -- the TM (pump) power content $P_{b}(0)$ space, for a
small TE initial power $P_{a}(0) = 0.03$ and zero angle $k_a = 0$,
and for initial linear  polarization.  The whiter region marks a
high gain limit (up to 60 times), while the black region marks a
low gain (0.7). As expected, the maximum gain is achieved when the
angle of the pump is near zero, and when the amplitude to the pump
is the highest. In Fig.~\ref{fig2} (point ``A'' in Fig.~
\ref{fig1}), we show the longest switching case (13 sites) which
is associated with a low power gain, $\sim$ 70$\%$. In
Fig.~\ref{fig3} (point ``B'' in Fig.~\ref{fig1}), we show a
switching for 4 sites with a medium range power gain of $\sim$
800$\%$. In Fig.~\ref{fig4} (point ``C'' in Fig.~\ref{fig1}), we
show an example of strongly suppressed switching with a huge power
gain of $\sim$ 6.000$\%$. It is very important to note that the TE
signal has a very low power, approximately 1.7 W. Another
important remark is that the TE signal has no initial kick. With
that, we want to be emphatic about the real possibilities of the
implementation of our concept. The idea suggested in
Fig.~\ref{fig1}, is that a very small unkicked signal arriving at
the middle of the array can, as a consequence of the polarization
mode instability, be amplified in {\em any amount} from $70\%$ to
$6000\%$, by judiciously choosing the pump (TM) input angle and
the pump power.

This effect is possible due to an interplay between two effects,
the polarization instability and the array discreteness. The lower
the input power, the lower is the PN barrier, and the larger is
the distance from the input waveguide that the soliton can
propagate. Before the soliton gets trapped, it loses part of its
energy in the form of radiation modes, adjusting its profile to
the discrete soliton mode corresponding to the output waveguide.
Thus, for a low power level (Fig.~\ref{fig2}), discrete solitons
will have more transversal mobility and will be trapped in a
waveguide located far from the input region. At high enough power
level, the switching will be at a near (Fig.~\ref{fig3}) or null
(Fig.~\ref{fig4}) distance from the input waveguide.

In conclusion, we have suggested a novel multi-port
amplifying/switching scheme, which uses the typical polarization
instability property of vector solitons as well as the PN concept
due to the system discreteness.

The authors acknowledge a support from a Comisi\'{o}n Nacional de
Investigaci\'{o}n Cient\'{i}fica y Tecnol\'{o}gica and Fondo
Nacional de Desarrollo Cientif\'{i}co y Tecnol\'{o}gico. R.A.
Vicencio's e-mail address is rodrigov@fisica.ciencias.uchile.cl

\newpage

\section*{List of Figure Captions}

Fig. 1. TE gain vs. initial kick $k_b$ and the normalized power
$P_{b}(0)$ of the TM mode ($k_a=0$, $P_a(0)=0.03$).

\begin{figure}[t]
\includegraphics{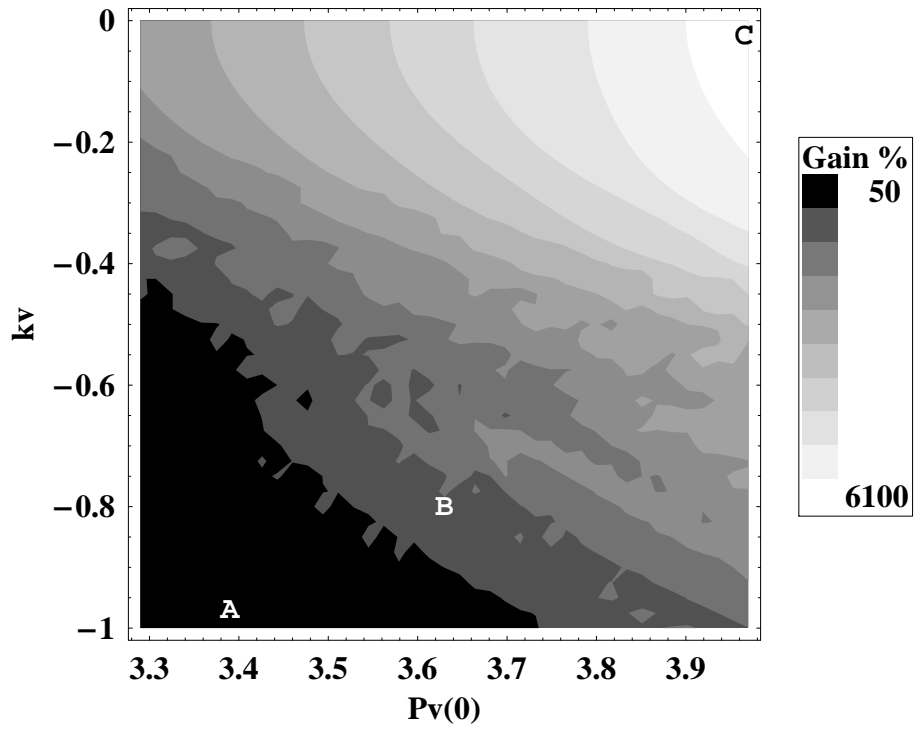}
\caption{TE gain vs. initial kick $k_b$ and the normalized power
$P_{b}(0)$ of the TM mode ($k_a=0$, $P_a(0)=0.03$).} \label{fig1}
\end{figure}

\begin{figure}[t]
\includegraphics{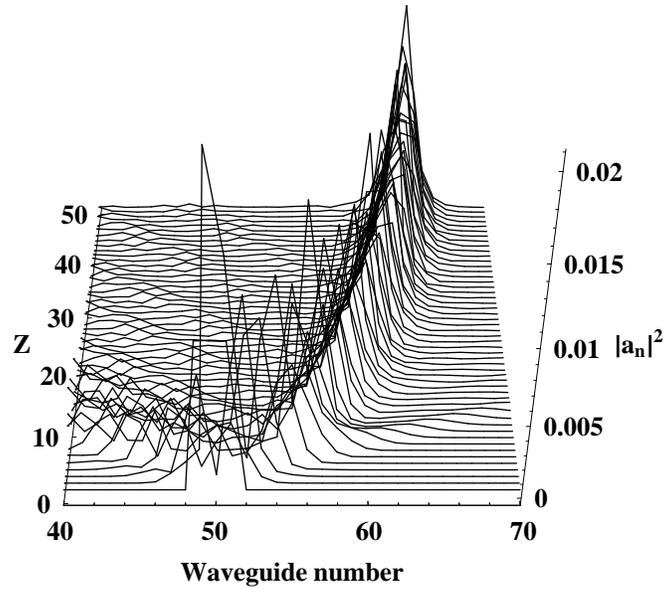}
\caption{Switching by 13 sites ($k_b = -1$, $P_b(0)=3.39$). The
gain is about $70\%$  (point ``A'' in Fig.~\ref{fig1}).}
\label{fig2}
\end{figure}

\begin{figure}[t]
\includegraphics{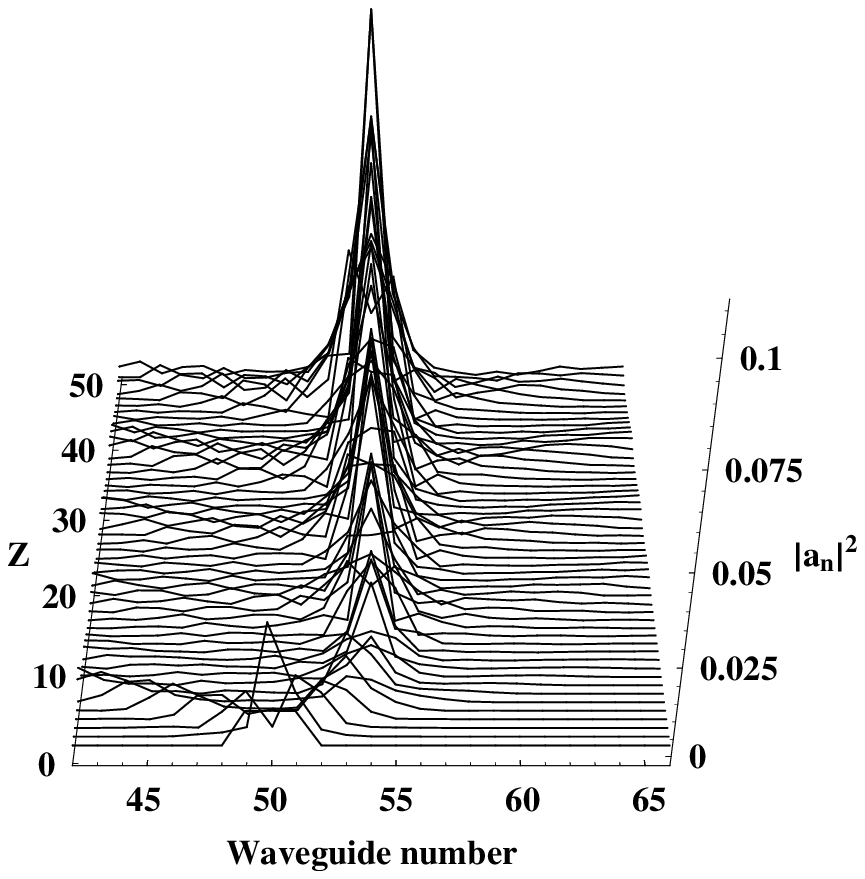}
\caption{Switching by 4 sites ($k_b = -0.8$, $P_b(0)=3.63$). The
gain is about $800\%$ (point ``B'' in Fig.~\ref{fig1}).}
\label{fig3}
\end{figure}

\begin{figure}[t]
\includegraphics{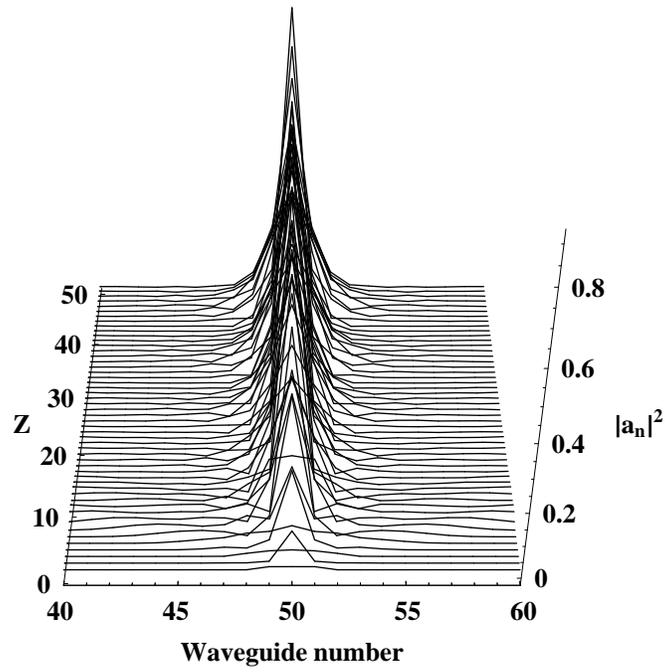}
\caption{Trapping without switching ($k_b = 0$, $P_b(0)=3.97$).
The gain is about $6.000\%$ (point ``C'' in Fig.~\ref{fig1}).}
\label{fig4}
\end{figure}

Fig. 2. Switching by 13 sites ($k_b = -1$, $P_b(0)=3.39$). The
gain is about $70\%$  (point ``A'' in Fig.~\ref{fig1}).

Fig. 3. Switching by 4 sites ($k_b = -0.8$, $P_b(0)=3.63$). The
gain is about $800\%$ (point ``B'' in Fig.~\ref{fig1}).

Fig. 4. Trapping without switching ($k_b = 0$, $P_b(0)=3.97$). The
gain is about $6.000\%$ (point ``C'' in Fig.~\ref{fig1}).

\newpage

\newpage

\newpage

\newpage

\end{document}